# Deterministic manipulation of multi-state polarization switching in multiferroic thin films

Chao Chen, Deyang Chen*, Peilian Li, Minghui Qin, Xubing Lu, Guofu Zhou, Xingsen Gao, Jun-Ming Liu

Chao Chen, Deyang Chen, Peilian Li, Minghui Qin, Xubing Lu, Xingsen Gao, Jun-Ming Liu
Institute for Advanced Materials, South China Academy of Advanced Optoelectronics, South China Normal University, Guangzhou 510006, China
E-mail: deyangchen@m.scnu.edu.cn (D. Chen)

Chao Chen, Deyang Chen, Guofu Zhou
Guangdong Provincial Key Laboratory of Optical Information Materials and Technology, South China Academy of Advanced Optoelectronics, South China Normal University, Guangzhou 510006, China

Jun-Ming Liu
Laboratory of Solid State Microstructures and Innovation Center of Advanced Microstructures, Nanjing University, Nanjing 210093, China

**Abstract:** Deterministically controllable multi-state polarizations in ferroelectric materials are promising for the application of next-generation non-volatile multi-state memory devices. However, the achievement of multi-state polarizations has been inhibited by the challenge of selective control of switching pathways. Here we report an approach to selectively control 71° ferroelastic and 180° ferroelectric switching paths by combining the out-of-plane electric field and in-plane trailing field in multiferroic BiFeO$_3$ thin films with periodically ordered 71° domain wall. Four-state polarization states can be deterministically achieved and reversibly controlled through precisely selecting different switching paths. Our studies reveal the ability to obtain multiple polarization states for the realization of multi-state memories and magnetoelectric coupling based devices.

# 1. Introduction

Electrical control of polarization states in ferroelectric materials plays a crucial role for the achievement of high-density non-volatile memory devices.[1-12] To overcome the inherent bi-stable polarization of ferroelectrics, numerous efforts have been made to obtain multiple polarization states for multi-state memory applications. For instance, tunable polarization values can be obtained by adjusting the fractions of switching polarization.[13-17] Moreover, the design of ferroelectric and dielectric multilayers or the manipulation of the coexisting structural variants (such as stripe a1/a2 domain structure in PbTiO$_3$-based thin films) can give rise to the formation of multiple polarization states.[6-11]

In multiferroic BiFeO$_3$ (BFO) thin films, the coexistence of ferroelectricity and antiferromagnetism as well as the magnetoelectric coupling effect provide new possibilities for the control of multi-state orders.[18-24] As a result of the [111]-oriented inherent polarization in rhombohedral BFO, three possible switching paths exist, including ferroelastic 71° and 109° switching, and ferroelectric 180° switching.[19] The sketches of 71° ferroelastic switching and 180° ferroelectric switching paths are schematically illustrated in Figure 1, revealing the importance to achieve ferroelastic switching paths for the realization of magnetoelectric devices as the 180° polarization reversal cannot directly change the orientation of the antiferromagnetic order.[19,22,23,25,26] However, it is a great challenge to obtain ferroelastic switching due to its high-energy domain state.[27-29]

Several significant studies have demonstrated the methods to manipulate the domain switching pathways in BFO thin films.[28,30,31] Baek et al. reported the control of 71° switching using monodomain BiFeO$_3$ islands.[30] Non-180° switching can also be obtained by controlling a scanning probe microscope (SPM) tip motion directions.[28] Very recently, selective control of 71° ferroelastic switching and 180° ferroelectric switching was proposed in the BFO film by trailing flexoelectric field.[31] However, these methods remain limited by the additional fabrication process,[30] non-reversible control of polarizations,[31] or insufficient investigation of the out-of-plane polarization switching.[28]

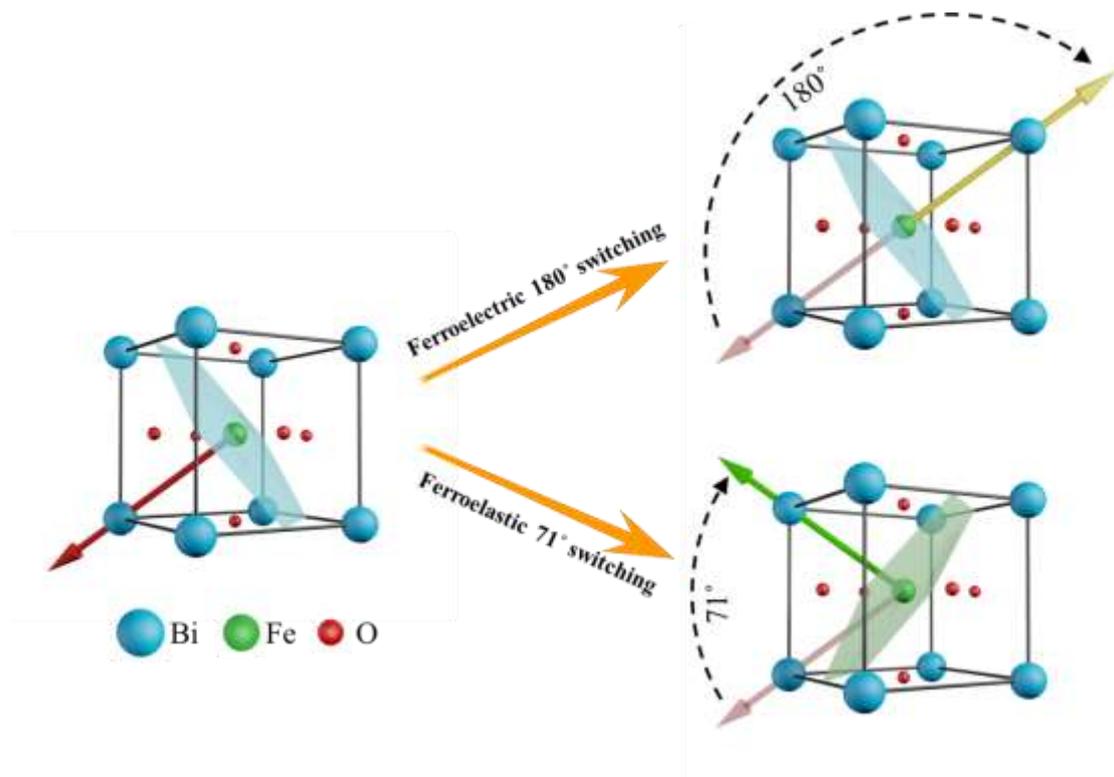

**Figure 1.** Sketches of 180° ferroelectric switching and 71° ferroelastic switching paths in BiFeO$_3$. The cubic frameworks, vectors and hexagonal planes represent the BiFeO$_3$ unit cells, ferroelectric polarizations and antiferromagnetic planes, respectively.

In this letter, we report the achievement and reversible control of multi-state polarization switching in multiferroic BFO thin films through the deterministic selection of ferroelectric and ferroelastic switching paths. The combination effects of out-of-plane electric field and in-plane trailing field were used to control the switching paths and polarization states in BFO thin films with periodical ordered 71° domain walls. By introducing the in-plane trailing field, we show that the ferroelastic 71° and ferroelectric180° switching paths can be selectively controlled without wrecking the stripe domain patterns. Furthermore, by the deterministic selection of the two switching paths, we demonstrate an approach to create four net polarization states with stripe domain configurations. Our findings open a pathway for the realization of multi-state memories as well as magnetoelectric devices based on multiferroic BFO.

## 2. Results and discussion

Arising from elastic energy relaxation and electrostatic equilibrium, periodic nanoscale arrays of 71° domain walls in BFO film with only two ferroelectric polarization variants can be produced by diminishing the variants through interface effect, screening effect and in-plane anisotropy of substrates.[18,28,32] Here, we successfully synthetized 71° stripe domain BFO thin films on $DyScO_3$ (DSO) (110) substrates with the conductive $SrRuO_3$ (SRO) as the bottom electrode layer. The BFO films exhibit atomically flat terrace-like morphology indicated by the Atomic Force Microscopy (AFM) image (Figure S1). The X-ray diffraction 2θ-ω curve (Figure S2), measured near the $(110)_o$ (where the subscript "O" denotes an orthorhombic index) peak of DSO substrate, shows only (002) diffraction peaks of BFO and SRO, confirming the epitaxial growth of high-quality BFO thin film. Moreover, the feature of Keissig fringe patterns indicate a quite flat surface, in consistent with the AFM data. The thickness of BFO film was calculated to be ~100 nm from periods of the fringe patterns.

Eight possible ferroelectric variants ($P_i^\pm$) in BFO are shown in Figure 2b. The OOP and IP PFM data of the as-grown BFO films are shown in Figure 2c and 2d, respectively. These PFM features, including uniform out-of-plane contrast and periodic stripe in-plane configuration, demonstrate that only two ferroelectric variants ($P_1^-$ and $P_2^-$) exist due to the monoclinic structure of DSO along $[1\bar{1}0]_o$ and the screening effect of SRO layer. This as-grown stripe domain structure is sketched in Figure 2e, in which the short-black arrows indicate the 71° stripe domain and the long red arrow reveals the net polarization ($P_{net}$) direction. Here we label this as-grown polarization state as Γ⁻. It is worth to mention that the 71° domain wall should be perpendicular to $P_{net}$ due to the electrostatic and ferroelastic effects.[18,33] Thus, the relationship of the 71° domain wall and ferroelectric polarization variants can be sketched in Figure 2f.

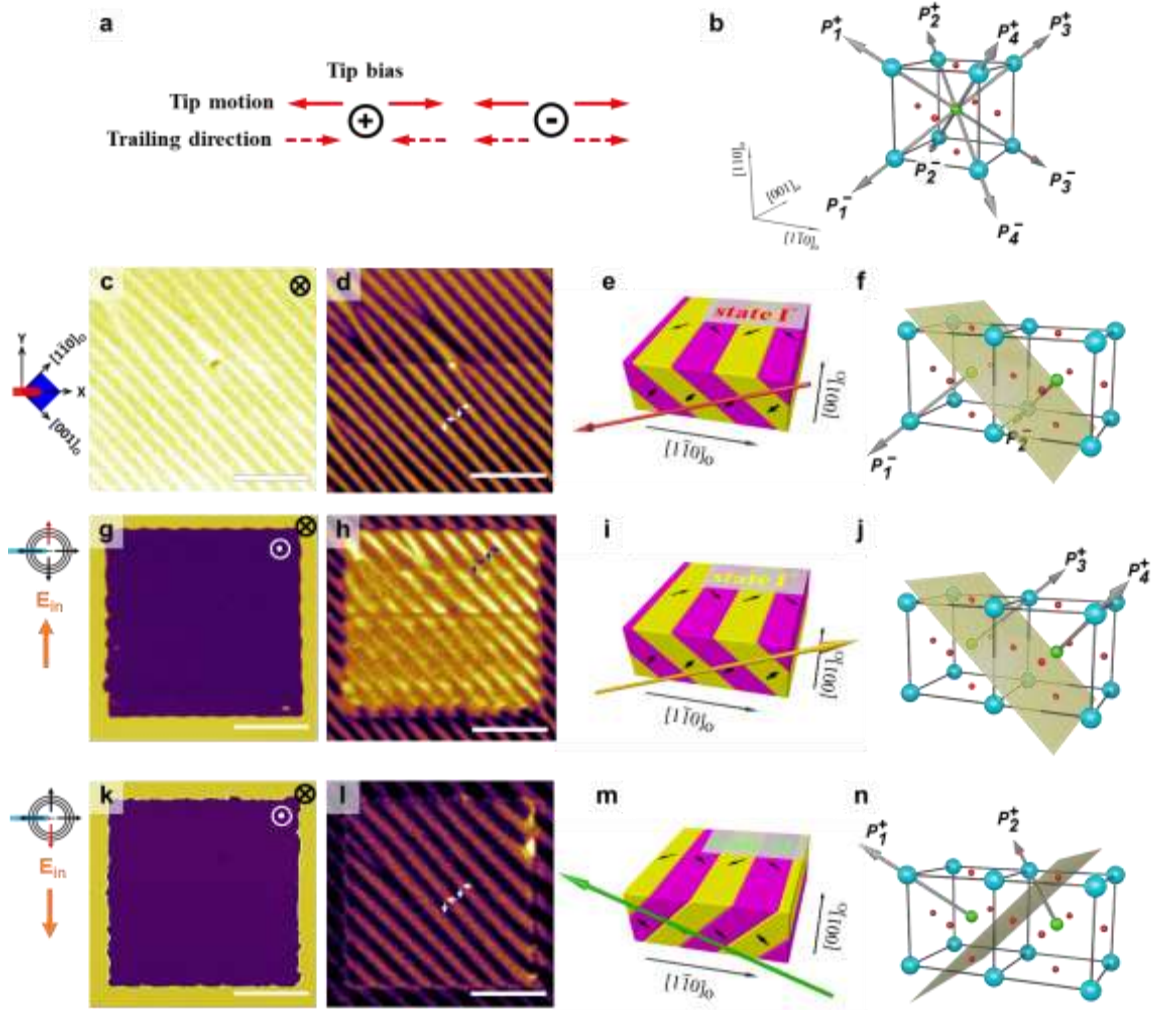

**Figure 2.** Deterministic selection of switching paths in 71° stripe domain BiFeO$_3$ films. (a) The sketches of the relationship between tip bias, motion and trailing field. (b) Unit cell of BiFeO$_3$ with 8 possible ferroelectric polarizations $P_i^\pm$ pointing along [111] crystal orientations. (c) Out-of-plane and (d) in-plane PFM images of the as-grown periodically ordered 71° domain walls in BiFeO$_3$ and corresponding (e) sketched domain structure and polarization state Γ⁻, and (f) the sketched as-grown polarization variants and 71° domain wall. (g-h) Corresponding PFM images and sketches after 180° ferroelectric switching. (k-n) Corresponding PFM images and sketches after 71° ferroelastic switching. Scale bar, 1 μm.

The trailing effect (also refers to the trailing field) can be equivalently regarded as an adjoint in-plane electric field generated by the moving SPM probe.[11,28,31,34-42] It was extensively employed for the fabrication of superdomains[11] and conductive domain walls,[31,34] as well as controllable magnetoelectric systems,[39,40] indicating its ability of tuning the in-plane polarization component. When the SPM probe with electrical bias relative to the grounding bottom electrode moves on the film surface, the direction of tip motion and the bias voltage co-determine the efficient in-plane trailing electric field direction, which is the basic law of trailing effect. As schematically illustrated in Figure 2a, upon the positive bias is utilized, tip motion direction is antiparallel with the trailing direction because the positive tip attracts negative tails of the polarization vectors. Otherwise, they are parallel. The tip motion is usually regarded as the slow scan direction in SPM, which is a net motion during scanning a square region.

Next, we turn to study the trailing effect on switching behaviors in multiferroic 71° stripe domain BFO thin films. In the first case, the BFO thin film was poled by -5 V voltage with trailing direction parallel to one of the variants in plane (X or Y, as shown in the sketched inset at the left side of Figure 2g), both the out-of-plane (Figure 2g) and in-plane (Figure 2h) PFM data exhibit obvious changes of the contrast, suggesting a conventional 180° ferroelectric switching path. The polarization state and domain structure after 180° switching are schematically illustrated in Figure 2i, labelled as state $\mathrm{I}^+$. As shown in Figure 2i and 2j, in contrast with the initial state $\mathrm{I}^-$, the 71° domain wall pattern is maintained while the net polarization direction is reversed and polarization variants are converted from $P_1^-$ and $P_2^-$ to $P_3^+$ and $P_4^+$, respectively, as a result of 180° switching.

However, upon the trailing direction reversed, i.e., along -X or -Y which is antiparallel to variants in plane, the switching behavior is quite different. Only the out-of-plane polarization (Figure 2k) is reversed while the in-plane polarization (Figure 2l) shows almost no contrast difference and sustains the periodic stripe feature. These results demonstrate the emergence of ferroelastic 71° switching behavior. The domain configuration after ferroelastic 71° switching is schematically reconstructed in Figure 2m and another polarization state is labelled as state $\mathrm{II}^+$, whose $P_{net}$ is different from state $\mathrm{I}^-$ (Figure 2e) and state $\mathrm{I}^+$(Figure 2i). The corresponding ferroelectric variant model of Figure 2k-2m is sketched in Figure 2n, which presents the 71° stripe domain pattern constituted by ferroelectric variants of $P_1^+$ and $P_2^+$.

Thus, we demonstrate the selective control of ferroelectric 180° switching and ferroelastic 71° switching paths using the effective in-plane trailing field (parallel or antiparallel to the variants). Besides, two more polarization states including state I$^+$ and state II$^+$ can be deterministically achieved through the control of switching paths based on the as-grown polarization state I$^-$. It is worth to mention that the 180° polarization reversal from state I$^-$ to state I$^+$ doesn't change the antiferromagnetic plane while the ferroelastic 71° switching from state I$^+$ to state II$^+$ does.

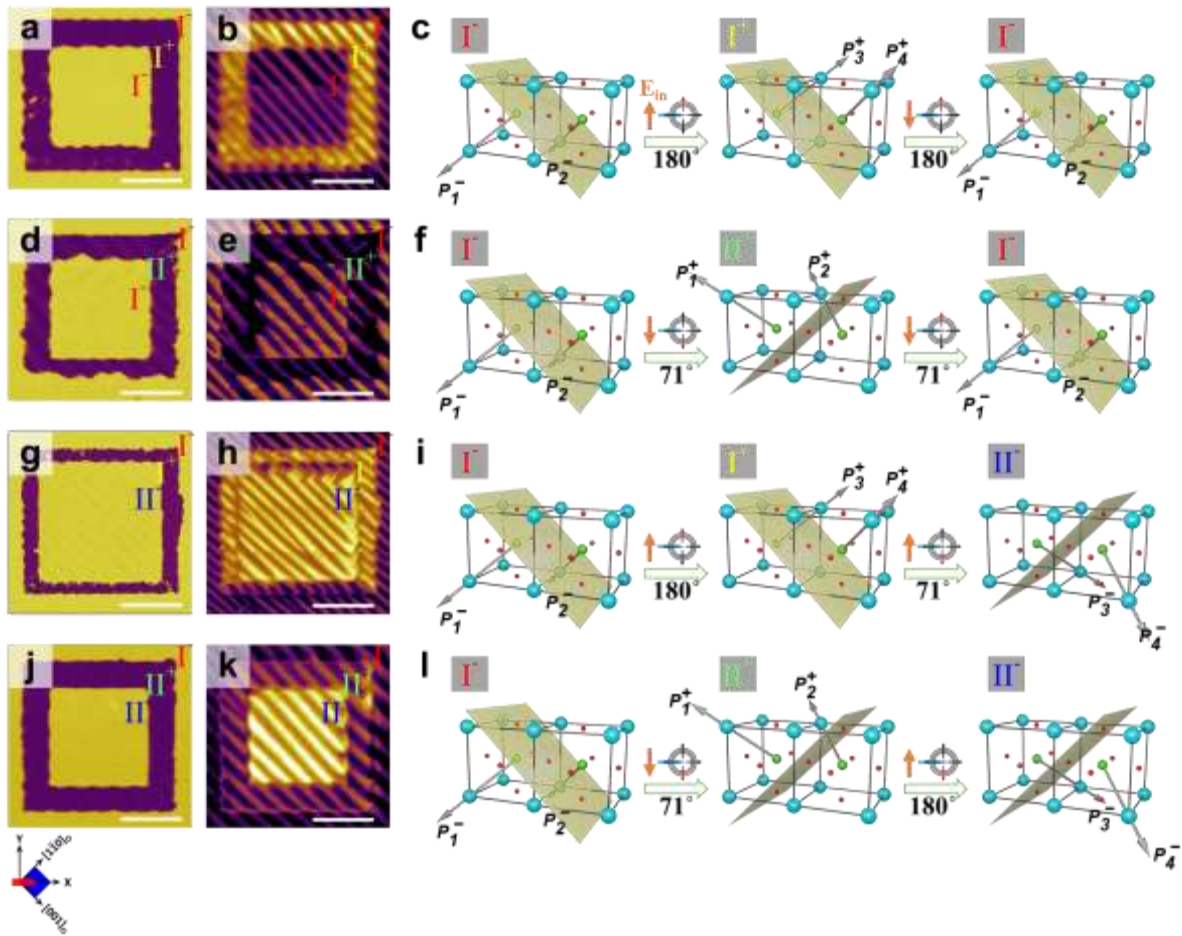

**Figure 3.** Reversible manipulation of multi-state polarizations through the deterministic control of the switching paths. (a-c) 180° ferroelectric switching path enabled reversible 180° polarization direction reversal from state I$^-$ to state I$^+$. (d-f) Pure 71° ferroelastic switching enabled the switching of polarization state I$^-$ and state II$^+$ back and forth. (g-l) Mixed switching paths of ferroelectric 180° switching and ferroelastic 71° switching paths giving rise to the deterministic achievement and reversible switching of four polarization states including state I$^-$, state I$^+$, state II$^+$ and state II$^+$. The right panel sketches (c, f, i, l) present the evolution of ferroelectric variants and domain wall architectures as well as the operation details of voltage bias and trailing field corresponding to out-of-plane (a, d, g, f) and in-plane (b, e, h, k) PFM data in left panel. Scale bar, 1 μm.

To verify the ability to reversible manipulation of multi-state polarizations through the deterministic control of the switching paths, a series of PFM experiments were further carried out based on the data shown in Figure 2. We applied +5 V voltage with various trailing directions to switch the polarization downward in the inner regions, as shown in Figure 3. PFM results after performing two switching paths are exhibited in left panel of Figure 3. The corresponding evolution of ferroelectric variants and domain wall architectures, as well as operation details including voltage bias and trailing field, are schematically illustrated in right panel of Figure 3. Reversible polarization state switching from state I$^-$ to state I$^+$ can be achieved by the conventional 180° ferroelectric switching path (Figure 3a-c). Pure 71° ferroelastic switching can be accurately selected as well (Figure 3d-f), leading to the switching of polarization state I$^-$ and state II$^+$ back and forth. Moreover, the mixed switching paths of ferroelectric 180° switching and ferroelastic 71° switching paths can be precisely selected to deterministically obtain and switch four polarization states including state I$^-$, state I$^+$, state II$^+$ and state II$^+$. In the meanwhile, the change of antiferromagnetic plane can be well-controlled through the selection of switching paths, which is critical for the realization of magnetoelectric coupling devices.

Thus, these experimental results and discussion allow us to give a full description of the deterministic manipulation of multi-state polarization switching by selective control of the switching paths in multiferroic BFO thin films with periodical ordered 71° stripe domain, as shown in Figure 4. Herein, the cycle arrows and the legends explain the details for realizing different switching paths and polarization states. By combining the out-of-plane electric field and in-plane trailing field, 71° ferroelastic switching and 180° ferroelectric switching paths can be deterministically selected, enabling the achievement of four polarization states (I$^+$, I$^-$, II$^+$ and II$^-$) and the reversible switching of these multi-state ferroelectric polarizations.

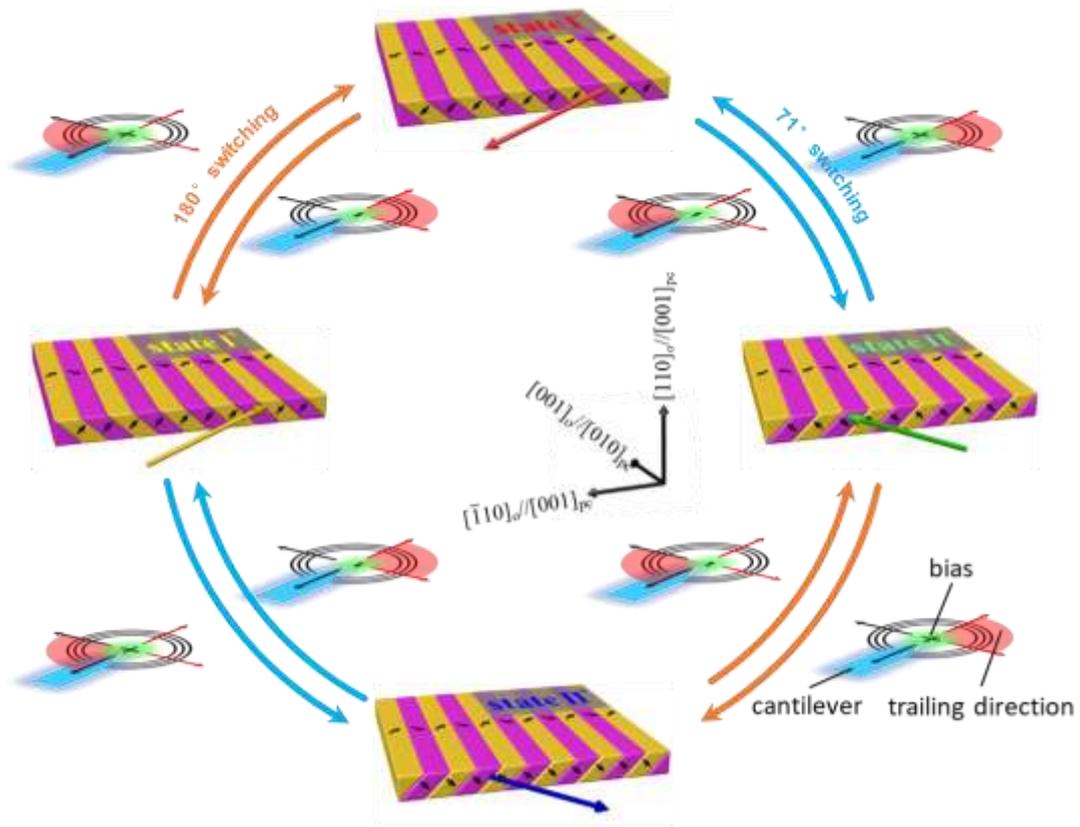

**Figure 4.** Full description of the deterministic manipulation of multi-state polarization switching by selective control of the switching paths in multiferroic BFO thin films with periodical ordered 71° domain walls. The cycle arrows and insets nearby show the corresponding operation details for the achievement of selective switching paths.

## 3. Conclusion

In summary, our studies overcome bi-stable ferroelectric polarization states by the selective control of 180° ferroelectric and 71° ferroelastic switching paths, giving rise to the deterministic manipulation of four-state polarization switching in multiferroic BFO thin films. These findings provide the platform for the realization of multi-state memory devices. Furthermore, this study would promote the application of BFO-based magnetoelectric devices owing to the achievement of the ferroelastic switching path to obtain the magnetoelectric coupling for electric field control of magnetism.[19,21-23,25,26,43]

## 4. Experimental Section

**Film Growth.** The stripe domain BFO thin films (thickness ~100 nm) and conductive SRO bottom electrode layers (thickness ~20 nm) were grown on $[110]_o$-oriented DSO substrates by pulsed laser deposition system with KrF excimer pulsed laser (where the subscript "O" denotes an orthorhombic index). The growth temperature of the substrates was kept at 700°C with the oxygen partial pressure of 15 Pa for both BFO and SRO layers.

**Characterization.** The crystal structures of the sample were examined by X-ray diffraction (XRD, PANalytical X'Pert PRO). The external electric field and trailing field in micron-size areas were all carried out with Pt-coated probes in scanning probe microscope (SPM, Cypher, Asylum Research). Furthermore, by using SPM, atomic force microscopy (AFM) and piezoresponse force microscopy (PFM) were performed to study the topography and domain architectures of BFO films. The PFM data in this work were the combination of the phase and amplitude data, which represent the orientation and intensity of the polarization, respectively. The data were synthetized following an equation given by $PFM = amplitude \times \sin(phase)$. In principle, there are three types of the contrasts in the IP PFM images, including dark, bright and neutral types, corresponding to the anti-clockwise or clockwise twist normal to cantilever and no twist. The dark and bright contrasts indicate that the in-plane components of the polarizations are perpendicular to the cantilever while the neutral one is parallel to it. During the measurement in SPM in this work, the relative position between the samples and probe were set up as the bearing mark shown at the left panel beside figure 2c, in which $[1\bar{1}0]_o$ and $[001]_o$ (subscript 'O' here denotes 'orthorhombic') axes refer to crystal orientation of DSO substrate. Therefore, the dark, bright and neutral contrast of the stripe patterns in IP PFM here indicate the -Y-, +Y- and -X-oriented IP polarizations, respectively.

**Supporting Information**

Supporting Information is available from the Wiley Online Library or from the author.

**Acknowledgments**

This work was supported by the National Natural Science Foundation of China (Grant No. 91963102). Authors also acknowledge the financial support of Guangdong Science and

Technology Project (Grant No. 2019A050510036), the Natural Science Foundation of Guangdong Province (Grant No. 2020A1515010736) and Guangdong Provincial Key Laboratory of Optical Information Materials and Technology (No. 2017B030301007). D.C. also acknowledges the support of 2022 International (Regional) Cooperation and Exchange Programs of SCNU.

**Conflict of Interest**

The authors declare no conflict of interest.

**Data Availability Statement**

The data that support the findings of this study are available from the corresponding author upon reason